\documentclass[%
 reprint,
 superscriptaddress,
 amsmath,amssymb,
 aps,
 pra,  
 longbibliography,
]{revtex4-2}

\usepackage{ulem}
\usepackage[utf8]{inputenc}
\usepackage[T1]{fontenc}
\usepackage[dvipsnames]{xcolor}
\usepackage{graphicx}% Include figure files
\usepackage{dcolumn}% Align table columns on decimal point
\usepackage{bm}% bold math
\usepackage{chemformula}
\graphicspath{ {./figures/} }
\usepackage[breaklinks,hypertexnames=false]{hyperref}
\hypersetup{colorlinks,linkcolor={magenta},citecolor={blue},urlcolor={blue}} 

\usepackage[capitalize]{cleveref}

\definecolor{crimson}{RGB}{255,102,255}
\definecolor{mahogany}{RGB}{192,64,0}

\usepackage{sidecap,tikz}
\definecolor{lime}{HTML}{A6CE39}
\DeclareRobustCommand{\orcidicon}{\hspace{-1.5mm}
	\begin{tikzpicture}
		\draw[lime, fill=lime] (0,0) 
		circle [radius=0.16] 
		node[white] {{\fontfamily{qag}\selectfont \tiny \,ID}};
		\draw[white, fill=white] (-0.0525,0.095) 
		circle [radius=0.007];
	\end{tikzpicture}
	\hspace{-3mm}
}
\foreach \x in {A, ..., Z}{\expandafter\xdef\csname orcid\x\endcsname{\noexpand\href{https://orcid.org/\csname orcidauthor\x\endcsname}
		{\noexpand\orcidicon}}
}

\begin{document}

\preprint{APS/123-QED}

\title{Signatures of edge states in antiferromagnetic van der Waals Josephson junctions}

\newcommand{\fmc}{Departament of Condensed Matter Physics, Universidad Aut\'{o}noma de Madrid, 28049 Madrid, Spain}
\newcommand{\ftmc}{Department of Theoretical Condensed Matter Physics, Universidad Aut\'onoma de Madrid, 28049 Madrid, Spain}
\newcommand{\ifimac}{Condensed Matter Physics Center (IFIMAC), Universidad Aut\'onoma de Madrid, 28049 Madrid, Spain}
\newcommand{\inc}{Instituto Nicol\'as Cabrera, Universidad Aut\'onoma de Madrid, 28049 Madrid, Spain}
\newcommand{\icmm}{Instituto de Ciencia de Materiales de Madrid (ICMM), Consejo Superior de Investigaciones Científicas (CSIC), Sor Juana Inés de la Cruz 3, 28049 Madrid, Spain}
\newcommand{\ua}{Laboratorio de Transporte Cu\'{a}ntico, Unidad Asociada UAM/ICMM-CSIC, Madrid, Spain}
\newcommand{\konstanz}{Department of Physics, University of Konstanz, Universitätsstraße 10, 78464 Konstanz, Germany}

\author{Celia Gonz\'alez-S\'anchez\orcidC{}}
\affiliation{\fmc}
\affiliation{\ifimac}
\affiliation{\inc}

\author{Ignacio Sardinero\orcidI{}}
\affiliation{\ftmc}
\affiliation{\ifimac}
%\affiliation{\inc}

\author{Jorge Cuadra}
\affiliation{\fmc}
\affiliation{\ifimac}
\affiliation{\inc}

\author{Alfredo Spuri\orcidT{}}
\affiliation{\konstanz}

\author{Jos\'e A. Moreno}
\affiliation{\fmc}
\affiliation{\ifimac}

\author{Hermann Suderow\orcidH{}}
\affiliation{\fmc}
\affiliation{\ifimac}
\affiliation{\inc}

\author{Elke Scheer\orcidS{}}
\affiliation{\konstanz}

\author{Pablo Burset\orcidP{}}
\affiliation{\ftmc}
\affiliation{\ifimac}
\affiliation{\inc}

\author{Angelo Di Bernardo\orcidA{}}
\affiliation{\konstanz}

\author{Rub\'en Seoane Souto\orcidR{}}
\affiliation{\icmm}
\affiliation{\ua}

\author{Eduardo J. H. Lee\orcidE{}}
\email{eduardo.lee@uam.es}
\affiliation{\fmc}
\affiliation{\ifimac}
\affiliation{\inc}
\affiliation{\ua}

\date{\today}% It is always \today, today,

\begin{abstract}

The combination of superconductivity and magnetic textures leads to unconventional superconducting phenomena, including new correlated and topological phases. Van der Waals (vdW) materials emerge as a versatile platform for exploring the interplay between these two competing orders. Here, we report on individual NbSe$_2$/NiPS$_3$/NbSe$_2$ Josephson junctions behaving as superconducting quantum interference devices (SQUIDs), which we attribute to the interplay between the superconductivity of NbSe$_2$ and the spin texture of the vdW antiferromagnetic insulator NiPS$_3$. This behavior persists for in-plane magnetic fields of at least 6 T and is the result of interference between separated transport channels. Microscopic modeling of the antiferromagnet insulator/superconductor (AFI/S) interface reveals the formation of localized states at the edges of the junction that can lead to channels that dominate transport. Our findings highlight AFI/S heterostructures as a platform for engineering novel superconducting phenomena, and establish a new route for lithographic-free SQUIDs that operate in high magnetic fields.

\end{abstract}

\maketitle

The interplay between superconductivity and magnetism leads to emergent phenomena that are absent in these ordering states alone, including the generation of spin-triplet Cooper pairs, topological states, and non-reciprocal supercurrents. Such phenomena are attracting a growing interest due to their potential for the development of 
superconducting devices with new functionalities and forms of control \cite{Linder2015, Sau_topology_2010}. Some of the key advances~\cite{Bergeret_RMP2018} have been achieved by studying superconductor/ferromagnet (S/F) and superconductor/ferromagnetic insulator (S/FI) heterostructures, e.g., the demonstration of spin-triplet pairing at S/F interfaces~\cite{DiBernardo2015a, DiBernardo2015b, Diesch2018}, millisecond-range spin-relaxation times of quasiparticles in S/FI systems~\cite{Parkin2010}, the measurement of long-range triplet supercurrents in S/F/S Josephson junctions~\cite{Khaire2010, Robinson2010}, and the generation of triplet supercurrents in non-equilibrium N/S/F/S/N systems (N being a normal metal) \cite{Jeon_2019}. By contrast, superconductor/antiferromagnet (S/AF) heterostructures have been studied to a much lower extent. Notably, AFs display a nearly zero net magnetization, which minimizes stray fields on the superconductor, and their spin texture can lead to very rich physics when combined with superconductivity \cite{jeon_long-range_2021, jeon_chiral_2023}. The synthetic spin-orbit field of an antiferromagnet has also been proposed to engineer topological superconductivity at S/AF interfaces \cite{lado_solitonicmodes_prr2020, lado_interactionTS_prr2021, sardinero2024}, which has motivated the study of magnetic domains as a source of magnetic anisotropy \cite{Morten_PRB2012, desjardins_synthetic_2019}. 

In this work, we push forward the exploration of hybrid S/AF heterostructures by studying transport in 
Josephson junctions with an antiferromagnetic insulator (AFI) barrier. 
More specifically, we investigate vertical van der Waals (vdW) junctions composed of a thin layer of the AFI \ch{NiPS3} embedded between two S electrodes made of few-layer \ch{NbSe2} flakes. Our work builds on previous studies on superconducting vdW devices, including Josephson junctions based on rotated \ch{NbSe2} flakes \cite{yabuki_supercurrent_2016, farrar_twistedNbSe2_2021}, graphene- \cite{kim_grapheneJJ_nanoletters, dvir_planargraphene_PRB, zalic_highBjunction_nanoletters} or \ch{Nb3Br8}-based \cite{wu_field-free_2022} junctions with \ch{NbSe2} electrodes, stacks of \ch{NbSe2} coupled to the helimagnet \ch{Cr1/3NbS2} \cite{Spuri2023}, and ferromagnetic vdW Josephson junctions, e.g., with F (\ch{Fe3GeTe}~\cite{hu_long-range-skin_2023}) or FI (\ch{Cr2Ge2Te6}) vdW weak links \cite{idzuchi_unconventional_2021, ai_van_2021, kang_van_2022}. 
Unlike previous works, the response in our Josephson junctions is dominated by localized channels. Supported by microscopic calculations, we ascribe the formation of these channels to the combination of the magnetic texture of \ch{NiPS3} and the superconductivity of \ch{NbSe2}. When such transmitting channels concentrate into two separated regions, our devices behave as atomic-scale superconducting quantum interference devices (SQUIDs) with oscillations visible up to in-plane magnetic fields of at least 6\,T, thus demonstrating a new lithographic-free route for the fabrication of SQUIDs that are compatible with high fields.

\begin{figure*}
    \centering
    \includegraphics[width=\linewidth]{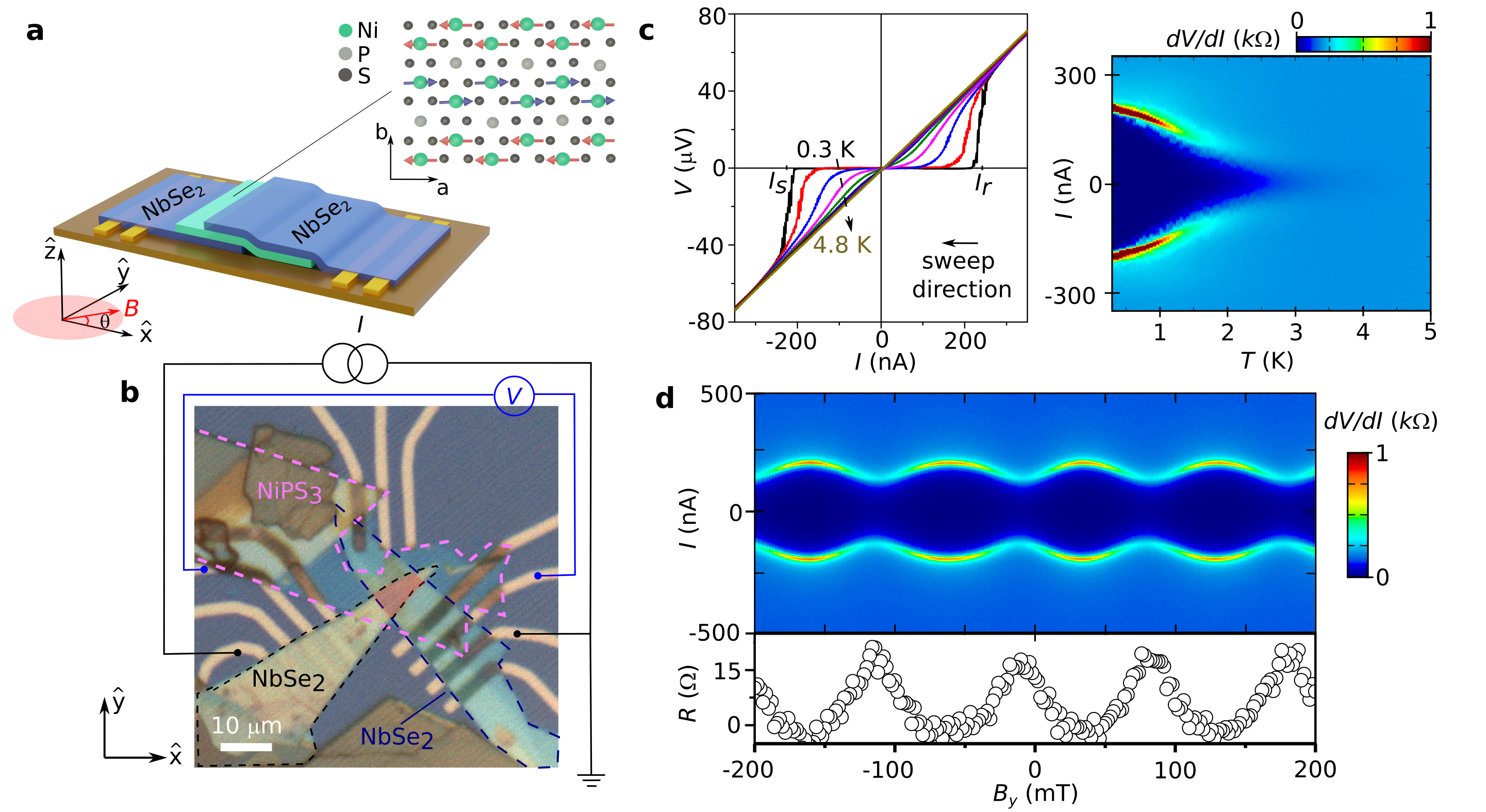}
    \caption{{\bf Antiferromagnetic van der Waals Josephson junctions}. 
    {\bf a}, Device schematics: a thin flake of the antiferromagnetic insulator \ch{NiPS3} (green) is embedded between two few-layer-thick superconducting \ch{NbSe2}  flakes (cyan). The yellow blocks represent Au bottom electrodes that are used as electrical contacts to the vdW heterostructure. The system is further subjected to an external in-plane magnetic field, $B$ (red arrow), which can be rotated in the $ab$-plane of the \ch{NiPS3} barrier. Top right: schematic representation of the spin texture in the $ab$-plane of \ch{NiPS3}. 
   {\bf b}, Optical micrograph of device A. Dashed lines define the edges of the \ch{NbSe2} (black and blue lines) and \ch{NiPS3} (magenta lines) flakes. A vertical Josephson junction is formed at the region in which the three flakes overlap (highlighted in red). 
   {\bf c}, Left panel, $V(I)$ curves taken at different temperatures for the device in (b), and right panel, differential resistance of the junction, $dV(I)/dI$, as a function of temperature. 
   {\bf d}, $dV(I)/dI$ as a function of the in-plane magnetic field applied along the $y$-axis, $B_y$, revealing characteristic SQUID oscillations. Bottom panel, zero-bias differential resistance, $R = dV(I = 0)/dI$, as a function of $B_y$.}
    \label{fig1}
\end{figure*}

\ch{NiPS3} is a layered antiferromagnet that, in its bulk form, has a monoclinic crystal structure with Ni atoms arranged in a honeycomb lattice. As schematically depicted in Fig.~\ref{fig1}a, the Ni atoms host localized magnetic moments that order antiferromagnetically below the N\'{e}el temperature, $T_N \approx 155$\,K. Previous studies have shown that the AF order in \ch{NiPS3} can be described, down to the bilayer limit \cite{kim_suppression-XXZ_2019}, by the XXZ model whereby the magnetic moments lie mostly in the \ch{NiPS3} layer $ab$ plane, with a small out-of-plane component in the $c$-axis and a slight anisotropy favoring alignment along the $a$-axis \cite{ joy_magnips_prb1992, wildes_nips_prb2015}. In the presence of an external applied magnetic field, $B$, \ch{NiPS3} magnetizes weakly due to spin canting, with a (weaker) stronger susceptibility along the ($a$-) $c$-axis. For higher in-plane magnetic fields of $B_{\parallel} \gtrsim 5$\,T, spin-flop transitions in the $ab$-plane of \ch{NiPS3} have also been reported \cite{basnet_spinflop_prm2021}.

\begin{figure*}
    \centering
    \includegraphics[width=\linewidth]{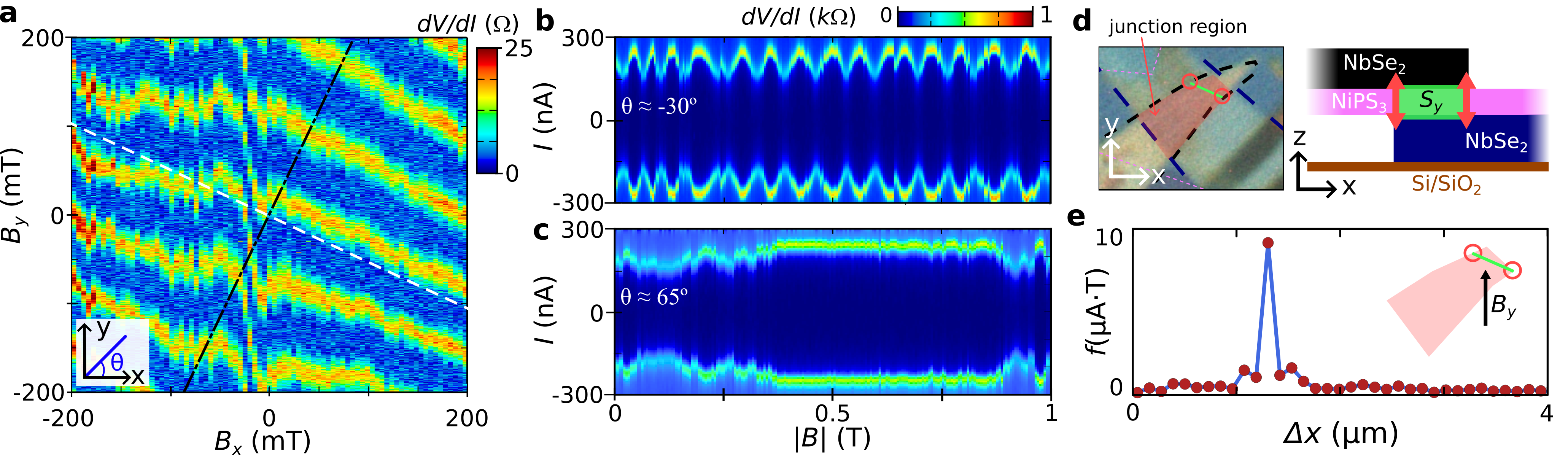}
    \caption{{\bf Angle-resolved supercurrent interference pattern.} {\bf a}, Zero-bias differential resistance as a function of the magnetic field components along the $\hat{x}$ and $\hat{y}$ axes, which lie in the $ab$-plane of the \ch{NiPS3} barrier and are perpendicular to the flow of the supercurrent. The white (black) dashed line underscores the direction parallel (perpendicular) to the diagonal lines formed by SQUID oscillations. {\bf b} and {\bf c}, $dV/dI$ as a function of the bias current and of an external magnetic field $B$ applied nearly perpendicular ($\theta \approx$ 65$^{\circ}$) and nearly parallel ($\theta \approx$ -30$^{\circ}$) to the diagonal lines in panel {\bf a}. {\bf d}, Schematic representation of the two localized channels, shown as red circles (arrows) in the left (right) panel. The green rectangle, $S_y$, in the right panel denotes the area perpendicular to the $\hat{y}$-axis. While the precise location of the channels is not known, their angle is inferred from the measurement in panel {\bf a}. 
    {\bf e}, Fourier transform of the $dV(I, B)/dI$ profile measured for $\theta = 90^{\circ}$, i.e., with $B$ along the $S_y$-axis (profile shown in the Supplemental Material). 
    } 
    \label{fig2}
\end{figure*}

To study the impact of a \ch{NiPS3} barrier on the Josephson transport of \ch{NbSe2}-based vdW junctions, we have prepared samples via the standard dry-transfer technique \cite{Castellanos-Gomez2014}, using few-layer \ch{NiPS3} and \ch{NbSe2} flakes (see Methods). The electronic transport properties of our devices have been measured using a four-terminal current-biased, $I$, setup and standard lock-in techniques (Fig.~\ref{fig1}b), from which the voltage drop, $V$, and the differential resistance, $dV/dI$, across the junction have been obtained simultaneously. We have also employed a vector magnet to study the impact of an applied magnetic field, $B$, which could be rotated within the $ab$-plane of the \ch{NiPS3} barrier. In the following, we focus on data acquired for device A. The description of the sample statistics can be found in Methods, and experimental data of additional devices in the Supplemental Material.  
Importantly, all samples that displayed Josephson effect have shown signatures of transport occurring via localized channels.

We start our experiment by characterizing the properties of bare \ch{NbSe2} flakes. 
From these measurements, we have determined  superconducting critical temperatures of $T_c^{\ch{NbSe2}} \approx 6-7$\,K, and critical currents of $I_c^{\ch{NbSe2}} \sim 0.1-1$\,mA at $T = 2$\,K, consistent with the behavior of few-layer \ch{NbSe2} 
(see Supplemental Material). We have then measured $V(I)$ and $dV(I)/dI$ across the vdW junction as a function of temperature, $T$. A non-hysteretic, dissipationless supercurrent branch can be clearly identified (see Fig.~\ref{fig1}c), displaying a switching current, $I_s$, of approximately 230 nA at $T = 0.3$\,K. The switching current gradually lowers with increasing temperature until it is fully suppressed at a critical temperature for the junction $T_{c}^{j} \gtrsim 4$\,K. We note here that $I_s$ is 3-4 orders of magnitude lower than $I_c^{\ch{NbSe2}}$, indicating that the supercurrent goes through the \ch{NiPS3} barrier. The normal-state resistance of the junction, $R_n \approx 200\Omega$, is taken as the zero-bias $dV/dI$ at $T = 5$\,K and allows us to estimate the  characteristic voltage, $V_c = I_sR_n \approx 46\,\mu$V, which is much smaller than the superconducting gap of \ch{NbSe2}, $\Delta \sim 1$\,meV \cite{Dvir_NatComm2018}. In addition, by estimating the junction area, $A$, to be $\approx$ 28\,$\mu$m$^2$, we obtain $\rho = R_nA \approx 5600\,\Omega \mu$m$^2$ and $J_s = I_s/A \approx 0.8$\,A/cm$^2$, which are consistent with a not very transparent junction. The found values are comparable with the ones reported in, for example, \ch{NbSe2} junctions with  6-10\,nm vdW FI barriers of \ch{Cr2Ge2Te6} \cite{kang_van_2022}. 

Despite the apparent low-transmission behavior of the junction, the temperature dependence of $I_s$ in our device is not well captured by the Ambegaokar-Baratoff relation~\cite{AB-formula, *AB-formula_err}. The trend of $I_s(T)$ is intricate and cannot be adjusted by simple models with only a few fitting parameters. The results of this analysis are reported 
in the Supplemental Material.

\begin{figure*}
    \centering
    \includegraphics[width=\textwidth]{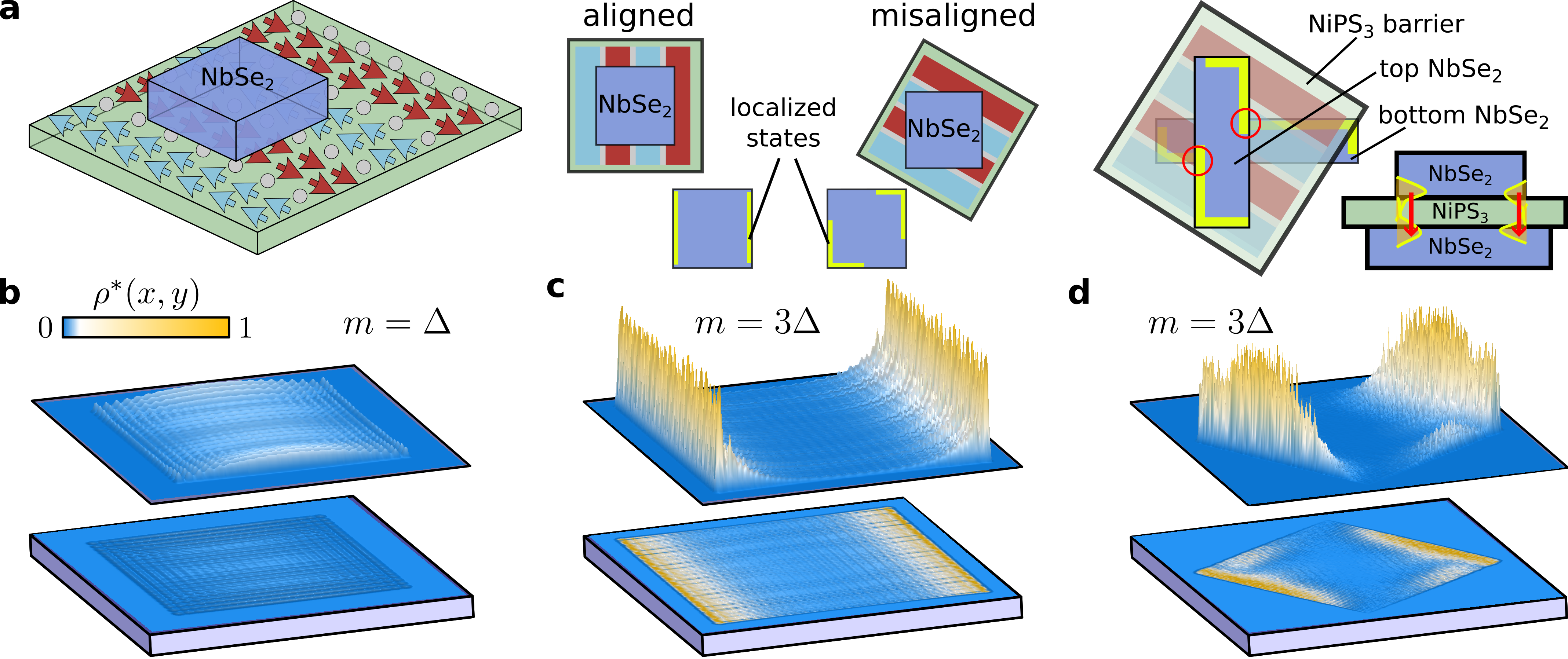}
    \caption{{\bf Theory model %describing the coupling between
    of the \ch{NbSe2}/\ch{NiPS3} interface.} {\bf a} Sketch of the S/AF heterostructure considered in the model. Right panel: Example of overlapping localized states at the top and bottom \ch{NbSe2}/\ch{NiPS3} interfaces, forming localized transport channels (red arrows). {\bf b-d} Weighted probability distribution $\rho^{*}(x,y)$ for $m = \Delta$ in {\bf b}, and for $m = 3\Delta$ in {\bf c} and {\bf d}, which features a misaligned S/AF interface as schematically depicted in the center panel of {\bf a}.}
    \label{fig3}
\end{figure*}

We next apply a magnetic field perpendicular to the junction. Fig.~\ref{fig1}d displays a measurement taken for $B \parallel \hat{y}$, where the supercurrent oscillates periodically with $B$. This behavior is different from that of a Fraunhofer pattern, characteristic of homogeneous supercurrent distributions. By contrast, the observed $I_s(B_y)$ dependence strongly resembles that of a SQUID, thus suggesting that the supercurrent in device A is predominantly carried by two spatially-separated channels. From the period of SQUID oscillations, $\Delta B_y \approx 100$\,mT, we estimate the effective area of the junction perpendicular to the field, $S_y = (d_\text{AF} + 2\lambda_L)W = \Phi_0/\Delta B_y \approx 2.06 \times 10^{-2}$\,$\mu$m$^2$, where $\Phi_0$ is the magnetic flux quantum, $W$ is the width of the junction perpendicular to $B_y$, 
$d_\text{AF}$ is the thickness of the AFI barrier, and $\lambda_L$ is the London penetration depth. 
By characterizing the thickness of similar \ch{NiPS3} flakes through optical contrast and atomic force microscopy, we obtain an upper bound  $d_\text{AF} \approx$ 10\,nm for our samples (see Supplemental Material). Considering $\lambda_L \sim$ 5\,nm \cite{yabuki_supercurrent_2016, kang_van_2022, wu_field-free_2022}, we estimate $W \gtrsim 1 \mu$m, consistent with the physical dimensions of our device (note that due to the shape of our vdW stack, $W$ varies from 0 to $\approx 5\mu$m, see Figs.~\ref{fig1}b and ~\ref{fig2}d). 
%Importantly, the well-defined SQUID oscillations with $B_y$ suggest the existence of the two dominant and well-localized transport channels in the junction.

To better understand the properties of the above channels, we next study the device properties as a function of the orientation of the applied magnetic field. Fig.\ref{fig2}a  shows the zero-bias differential resistance of the device measured as a function of $B_x$ and $B_y$. The $dV/dI$ peaks reflect the minima of $I_s(B)$, as also seen in Fig.~\ref{fig1}d. Therefore, they can be used to track the dependence of the SQUID oscillations with $B$. Two main features are observed in this measurement. First, the $dV/dI$ peaks form a series of approximately parallel diagonal lines in the $B_x$-$B_y$ plane (the dashed white line in Fig.~\ref{fig2}a is a guide to the eye). In addition, we identify a single phase slip event taking place near $B_x = 0$, which causes the diagonal lines to shift by exactly one magnetic flux quantum. The SQUID pattern above lacks oscillations when $B$ is applied parallel to the diagonal lines ($\theta \approx -25^{\circ}$) indicating that, for this angle, the SQUID formed in our device is not threaded by a magnetic flux. This is further confirmed by measurements of $dV(I)/dI$ taken as a function of $B$ perpendicular ($\theta \approx$ 65$^{\circ}$) or nearly parallel ($\theta \approx$ -30$^{\circ}$) to the diagonal lines, as shown in Figs.~\ref{fig2}b and \ref{fig2}c, respectively.

Further insight into our device is gained by means of supercurrent spatial density profiles across the junction~\cite{Dynes-Fulton_1971}, obtained by performing Fourier transforms of the $dV(I, B)/dI$ measurements (for more details, see the Supplemental Material). In Fig.~\ref{fig2}e, we plot the Fourier transform of a measurement taken for $B \parallel \hat{y}$ (Fig.~\ref{fig1}d and Fig. S7a), which displays an inhomogeneous supercurrent distribution featuring unequal supercurrent peaks separated by a distance $\Delta x\sim 1.1$ $\mu$m, where again we have fixed $(d_\text{AF} + 2\lambda_L) = $\,20\,nm. By contrast, the spatial distribution obtained for the measurement taken with $B$ parallel to the diagonal lines (Fig.~\ref{fig2}c) is almost featureless since the dependence of $dV/dI$ on $B$ is very weak (see Supplemental Material). Therefore, the Josephson transport in our device is dominated by two sets of localized channels, which our model below predicts to appear at edges of the S/AFI/S stack. As schematically depicted in Fig.~\ref{fig2}d, these channels do not necessarily localize at the most distant edges of the junction. Note that their position in the schematic (red circles) is only illustrative, as this information cannot be retrieved from the experimental data. Their orientation with respect to the $\hat{x}$ and $\hat{y}$ axes, on the other hand, is inferred from the measurements 
in Fig.~\ref{fig2}a. The position of the localized channels changes slightly with the magnetic field, which results in the diagonal lines in Fig.~\ref{fig2}a 
slightly changing slope and distance in the $B_x-B_y$ plane.

\begin{figure*}
    \centering
    \includegraphics{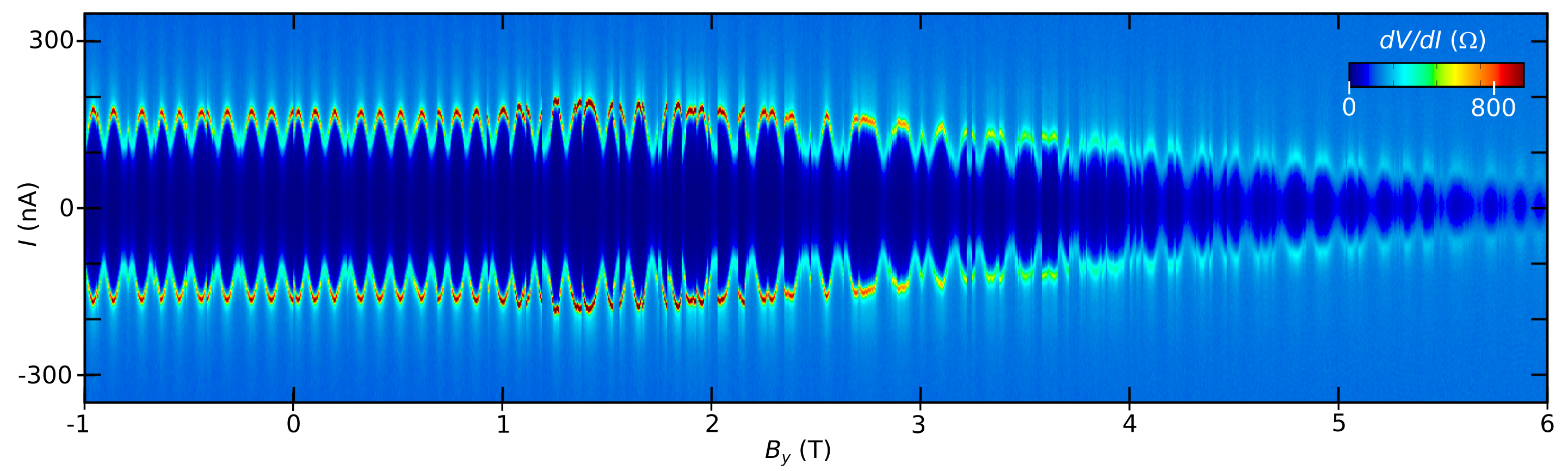}
    \caption{{\bf High magnetic field supercurrent quantum interference.} $dV(I, B)/dI$ measurements, demonstrating the persistence of SQUID oscillations up to at least 6\,T. 
    }
    \label{fig4}
\end{figure*}

We now present our theoretical model, which demonstrates that localized states emerge at NbSe$_2$/NiPS$_3$ interfaces due to the interplay between superconducting pairing and the antiferromagnetic order of NiPS$_3$\footnote{Our model supports localized states with trivial and topological origins, but the nature of the measured channels remains a subject of future research.}. 
Using a tight-binding approach, we simulate the S/AFI interface and compute the local density of states on a finite planar surface where superconductivity and antiferromagnetism coexist. 
Our model reproduces the NiPS$_3$ spin alignment in a square lattice with alternating spins encompassing a smaller region with superconducting pairing (left panel of Fig.~\ref{fig3}a). The resulting behavior is governed by the ratio of the onsite magnetization $m$ and the local pairing amplitude $\Delta$, assumed isotropic as in conventional s-wave superconductors.
 
Antiferromagnetic order implies zero net magnetization in the bulk, but local symmetry breaking at AFI/S boundaries leads to uncompensated moments. 
This localized magnetization facilitates the formation of low-energy subgap states near the Fermi level, $\epsilon_F$, particularly when $m \gtrsim \Delta$. 
To capture their spatial and energetic characteristics, we define a weighted probability distribution $
\rho^*(x,z) = \sum_{j>0} |\psi_j(x,z)|^2 / (\epsilon_j - \epsilon_F)^2
$. 
For $m/\Delta \lesssim 1$ $\rho^*$ is flat, indicating delocalized states (Fig.~\ref{fig3}b). For $m/\Delta \gtrsim 1$, distinct low-energy edge states emerge at regions of net boundary magnetization, as shown in Figs.~\ref{fig3}c and \ref{fig3}d. 
Due to different flake alignments in real samples, the two S/AFI interfaces in a Josephson junction likely exhibit different edge orientations, allowing localized states to overlap and form coherent transport channels (see rightmost panel of Fig.~\ref{fig3}a).

We have so far neglected the magnetization change of \ch{NiPS3} with $B$, as our main findings up to this point were made at low fields ($|B| \leq 0.2$ T). To address possible magnetization effects, we refer to $dV(I, B)/dI$ measurements where the field reaches 1 T (Figs.~\ref{fig2}b and ~\ref{fig2}c; see Fig. S7 for other angles). We have observed that, except for $\theta = 90^{\circ}$ (i.e., $B \parallel \hat{y}$), the regularity and periodicity of SQUID oscillations depend on the sweep direction of $B$ (see Fig. S8). From this observation, we hypothesize that $B_y$ is approximately aligned to the $a$-axis of \ch{NiPS3} where the magnetic susceptibility is the weakest \cite{joy_magnips_prb1992, wildes_nips_prb2015, basnet_spinflop_prm2021}. In this manner, the dependence of the supercurrent interference pattern on the sweep direction could be related to spin canting. We note, however, that a more detailed study of magnetization effects on the Josephson transport of \ch{NiPS3} junctions is beyond the scope of this work.

Finally, in Fig.~\ref{fig4}, we investigate the robustness of the SQUID interference for even higher magnetic fields. To this end, we apply a magnetic field in the range of $-1$ to 6\,T parallel to $\hat{y}$, to minimize changes on the \ch{NiPS3} magnetization. We observe that the SQUID oscillations persist up to the highest applied field, surpassing the maximum field achieved in previously reported SQUIDs \cite{zalic_highBjunction_nanoletters}. The oscillations are most regular (i.e., the periodicity and amplitude of the supercurrent remain approximately constant) for $|B| \lesssim 1$ T. Curiously, for slightly higher $B$, $I_s$ first increases with applied field before the amplitude starts to gradually decrease. Moreover, abrupt jumps can be observed, mostly in the range $B \gtrsim 1$. These jumps could possibly reflect the rearrangement of spins in the \ch{NiPS3} layer or vortices entering the vicinity of the junction.
Overall, the observed behavior demonstrates that the two localized channels are robust up to high magnetic fields. The changes in the supercurrent interference pattern for $B \gtrsim 1$ T, on the other hand, also suggest that the magnetization of the antiferromagnet becomes important in this regime, an effect that warrants further study.
It is worth noting that other examples of SQUID devices based on vdW materials have been reported \cite{Kim2021, Farrar2021}, although they require patterning into a loop geometry. In our device, the SQUID behaviour is obtained from a single junction and without any needs for patterning steps to define a SQUID loop.

To conclude, we have implemented antiferromagnetic van der Waals Josephson junctions with \ch{NiPS3} barriers. We have found that transport is dominated by localized channels located at $n=2$ regions at the junction edges. While the discussion in the main text has focused on one device, in the Supplemental Material we include data of additional samples with behaviors consistent with $n = 2$, similar to device A, and $n = 1$. Unlike previous works that have reported SQUID-like oscillations due to boundary states intrinsically present in the weak link, e.g., in topological \cite{hart_inducedHgTe_2014, Li_BiNW_2014, Murani_ballisticBi_2017, Chu_broad_Cd3As2_2023} or unconventional \cite{Le_NiTe2_2024} materials, or in domain walls in twisted bilayer graphene \cite{Barrier_SC_QHE_one-dimensional_graphene_2024}, the formation of localized channels in this work is attributed 
to the interplay between the magnetic texture of \ch{NiPS3} and the superconducting correlations of the electrodes. Our work thus contributes to the exploration of hybrid systems that combine superconducting correlations, spin texture and spin-orbit coupling, and establishes the vast, yet largely unexplored potential that superconductor/antiferromagnet hybrids based on van der Waals materials for the discovery of novel physical effects. Additionally, our demonstration of a SQUID device that is based on a single vdW junction and does not require  any patterning steps for the definition of its geometry can have implications on the realization of SQUID magnetometers with high scalability and on the integration of SQUID devices into superconducting circuits.

\bigskip
\textbf{\large{Methods}}

\textbf{Device fabrication: } The \ch{NbSe2/NiPS3/NbSe2} heterostructures in this work were assembled using standard dry-transfer techniques \cite{Castellanos-Gomez2014}. In short, flakes of \ch{NbSe2} and \ch{NiPS3} (HQ Graphene Inc.) were mechanically exfoliated from their respective bulk crystals onto PDMS stamps. Optical contrast was used to provide an estimation of the flake thickness, thus allowing us to select few-layer flakes for the preparation of heterostructures. By employing an optical microscope and a micro-manipulator, the selected flakes were aligned and brought into contact with bottom Cr/Au (2.5\,nm/30\,nm) electrodes, which were pre-fabricated on  undoped \ch{Si/SiO2} substrates. The PDMS stamp was then slowly peeled off, leaving the flake deposited on the substrate. The cleaved surfaces of the crystals create a contact by means of vdW interaction. To prevent sample degradation under ambient conditions, we have capped our \ch{NbSe2/NiPS3/NbSe2} heterostructures with a top hexagonal boron nitride (hBN) layer. The entire process above is carried out at room temperature. Samples were either prepared inside a glovebox filled with nitrogen gas, including device A discussed in the main text, or in air. Data for additional samples can be found in the Supplemental Material.  

\textbf{Sample statistics:} In total, we have measured 7 devices prepared in nitrogen atmosphere, from which two showed Josephson coupling (device A and device B; data of the latter is shown in the Supplemental Material). The others displayed higher resistances and no supercurrent, indicative of a thicker \ch{NiPS3} flake or of a poor engagement between the different vdW flakes \cite{Hu2025}. We have additionally measured 8 samples prepared in air, with three displaying supercurrents. However, most of these devices showed instabilities manifested as abrupt jumps in the transport, possibly related to degradation in air. A summary of this data is also shown in the Supplemental Material. 

\textbf{Electrical measurements:} Our experiment was carried out using two different cryogenic systems: a $^3$He insert with a base temperature of 250 mK, and a dilution refrigerator with a base temperature of 10\,mK. The former (latter) was employed for taking the measurements shown in Fig.~\ref{fig1} (Figs.~\ref{fig2} and \ref{fig4}). The DC wiring of systems included $\pi$ filters at room temperature,  followed by low-temperature RC filters with a cut-off frequency of 10 kHz. For the lines of the dilution refrigerator, we additionally installed low-pass filters with cut-off frequencies of 80 MHz, 1450\,MHz and 5000\,MHz at the level of the mixing chamber. 

We have employed current-bias, $I$, four-terminal measurements and standard lock-in techniques to characterize the transport response of our devices. In short, a current source (Standford Research Systems, CS580) was connected to the outer electrodes of the devices, while the inner pair of electrodes were used to measure the corresponding voltage drop, $V$. A small, low-frequency (amplitude = 10 nA, $f = 37$ Hz) AC signal supplied by a lock-in amplifier (Zurich Instruments, MFLI) was superimposed to the DC current, thus allowing to acquire the differential resistance, $dV/dI$, of the device simultaneously.

In order to take all the measurements, device A was thermally cycled five times within a period of a year. Only in the last cool-down the device showed signs of degradation, and eventually stopped working.

\textbf{Shift of SQUID oscillations from zero-field:} Note that the maximum $I_s$ in Fig.~\ref{fig1}d is displaced from $B_y = 0$ by $\delta \approx$ 35\,mT. We have found that $\delta$ is sensitive to the history of magnetic field sweeps in a non-monotonic way, possibly reflecting changes in the magnetization of the \ch{NiPS3} barrier.

\bigskip
\textbf{\large{Acknowledgments}}
\bigskip

We acknowledge funding by EU through the European Research Council (ERC) Starting Grant agreement 716559 (TOPOQDot), by the Spanish AEI through Grants No.~PID2020-117671GB-I00, PID2022-140552NA-I00, PID2023-150224NB-I00, TED2021-130292B-C41, PID2020-117992GA-I00, PID2020-114880GB-I00 and No.~CNS2022-135950, by Spanish CM ``Talento Program'' project No.~2019-T1/IND-14088, No.~2022-T1/IND-24070, and No.~2023-5A/IND-28927, through the ``Mar\'{\i}a de Maeztu'' Programme for Units of Excellence in R\&D (CEX2018-000805-M) and the ``Ram\'{o}n y Cajal'' programme grant RYC-2015-17973 as well as by the Deutsche Forschungsgemeinschaft (DFG- German Research Foundation) through project ID~443404566. 
The authors acknowledge Diego Alonso for his contribution to this study through the Atomic Force Microscopy measurements, which played a significant role in the characterization of the NiPS$_3$ flakes and Jesús Nevado for his help with the Raman Spectroscopy measurements. The authors also wish to thank Hadar Steinberg for useful discussions.

\bibliography{NiPS3_paper}

\end{document}